\documentclass[a4paper,11pt]{article}
\usepackage{pos}

\title{Beyond the Standard Model with BlackHawk v2.0}

\author*[a]{J\'er\'emy Auffinger}
\author[a,b]{Alexandre Arbey}

\affiliation[a]{Univ Lyon, Univ Claude Bernard Lyon 1,\\ CNRS/IN2P3, IP2I Lyon, UMR 5822, F-69622, Villeurbanne, France}

\affiliation[b]{Theoretical Physics Department, CERN, CH-1211 Geneva 23, Switzerland}
\affiliation[c]{Institut Universitaire de France (IUF), 103 boulevard Saint-Michel, 75005 Paris, France}

\emailAdd{j.auffinger@ipnl.in2p3.fr}
\emailAdd{alexandre.arbey@ens-lyon.fr}

\abstract{We present the new version of \texttt{BlackHawk v2.0}. \texttt{BlackHawk} is a public code designed to compute the Hawking radiation (HR) spectra of (primordial) black holes (PBHs). In the version \texttt{2.0}, we have added several non-standard BH metrics: charged, higher dimensional and polymerized black holes, in addition to the usual rotating (Kerr) BHs. \texttt{BlackHawk} also embeds some additional scripts and numerical tables that can prove useful in \textit{e.g.}~dark matter (DM) studies. We will describe these new features and provide some examples of the capabilities of the code. A tutorial for \texttt{BlackHawk} is available on the TOOLS2021 website: \url{https://indico.cern.ch/event/1076291/contributions/4609967/}}

\FullConference{%
  Computational Tools for High Energy Physics and Cosmology (CompTools2021)\\
  22-26 November 2021\\
  Institut de Physique des 2 Infinis (IP2I), Lyon, France
}

\newcommand{\mrm}[1]{_{\rm #1}}
\renewcommand{\d}{{\rm d}}

\begin{document}
\maketitle

\section{Introduction}

Since Hawking first demonstrated that black holes (BHs) evaporate by emitting a radiation close to the thermal radiation of a black body~\cite{Hawking:1974rv,Hawking:1975vcx}, this phenomenon has been extensively studied. One of the main outcome of HR is the possibility that small BHs (\textit{i.e.}~with mass $M \ll M_\odot$), formed just after the end of the universe expansion, denoted as PBHs, may emit or have emitted radiation that could be observable today or that could have left an imprint in cosmology. This leads to a number of constraints on the PBH abundance, depending on their mass $M$, dimensionless spin $a^*$, or the extended distribution of both these parameters. For recent reviews on PBH formation and constraints, see Refs.~\cite{Carr:2020gox,Green:2020jor}. When considering PBHs with mass $M \gtrsim M\mrm{eva}$ where $M\mrm{eva} \simeq 5\times 10^{14}\,$g is the mass of (Schwarzschild) PBHs evaporating just today if formed at the beginning of the universe, these constraints are given as the fraction of dark matter (DM) $f\mrm{PBH}$ that the PBHs represent today. For lighter BHs with mass $M \lesssim M\mrm{eva}$, these constraints are given as the maximum possible fraction of the universe $\beta$ collapsed into PBHs at their formation. These BHs cannot represent a significant fraction of DM today since they have evaporated away. However, if evaporation stops at some point, leaving Planck scale remnants, these may contribute to DM~\cite{MacGibbon:1987my,Barrow:1992hq}.

To determine the PBH evaporation constraints, one compares the HR signals to astrophysical or cosmological observations such as \textit{e.g.}~the extragalactic X/$\gamma$-ray background (EGXB) or the $\Delta N\mrm{eff}$ limits respectively. Noteworthy, HR is a purely gravitational phenomenon. As such, BHs would radiate any dark sector particle in addition to the Standard Model (SM) spectrum, such as supersymmetric states~\cite{Baker:2021btk}, DM or dark radiation (DR), \textit{e.g.}~\cite{Masina:2020xhk,Auffinger:2020afu,Masina:2021zpu,Arbey:2021ysg}.

It is thus of utmost importance to be able to predict precisely the HR spectra of SM and beyond SM (BSM) particles. There was thus a need for an automatic program to compute the precise spectra for any BH mass and spin, with a wide choice of parameters that allow for any kind of HR study. This was the context of creation of the public code \texttt{BlackHawk}~\cite{Arbey:2019mbc}. Since its first release, the code has received some modifications and has reached version \texttt{v2.1}~\cite{Arbey:2021mbl}. The code is publicly available on HEPForge:
\begin{center}
	\url{https://blackhawk.hepforge.org/}
\end{center}
\texttt{BlackHawk} is used by many groups from very different domains of astrophysics and cosmology to perform striking studies, a complete list of which is given on the \texttt{BlackHawk} website. \texttt{BlackHawk v2.1} includes some primordial new features linked to the physics described above: dark sector emission (Section~\ref{sec:DS_emission}), spin $3/2$ greybody factors (Section~\ref{sec:32}), BSM BH metrics (Section~\ref{sec:BSMmetrics}), low energy hadronization (Section~\ref{sec:Hazma}). The updated version of the code is available on the \texttt{BlackHawk} website mentioned above, and an updated version of the manual is available on the arXiv~\cite{Arbey:2019mbc} (v3). All the installation and run procedures, as well as the complete set of parameters and routines are described in the latter, while we focus here on the new features only (for a complete description see the release note~\cite{Arbey:2021mbl}).

\section{New features of BlackHawk}

\subsection{Dark sector emission}
\label{sec:DS_emission}

\begin{figure}
	\centering
	\includegraphics[width = 0.49\textwidth]{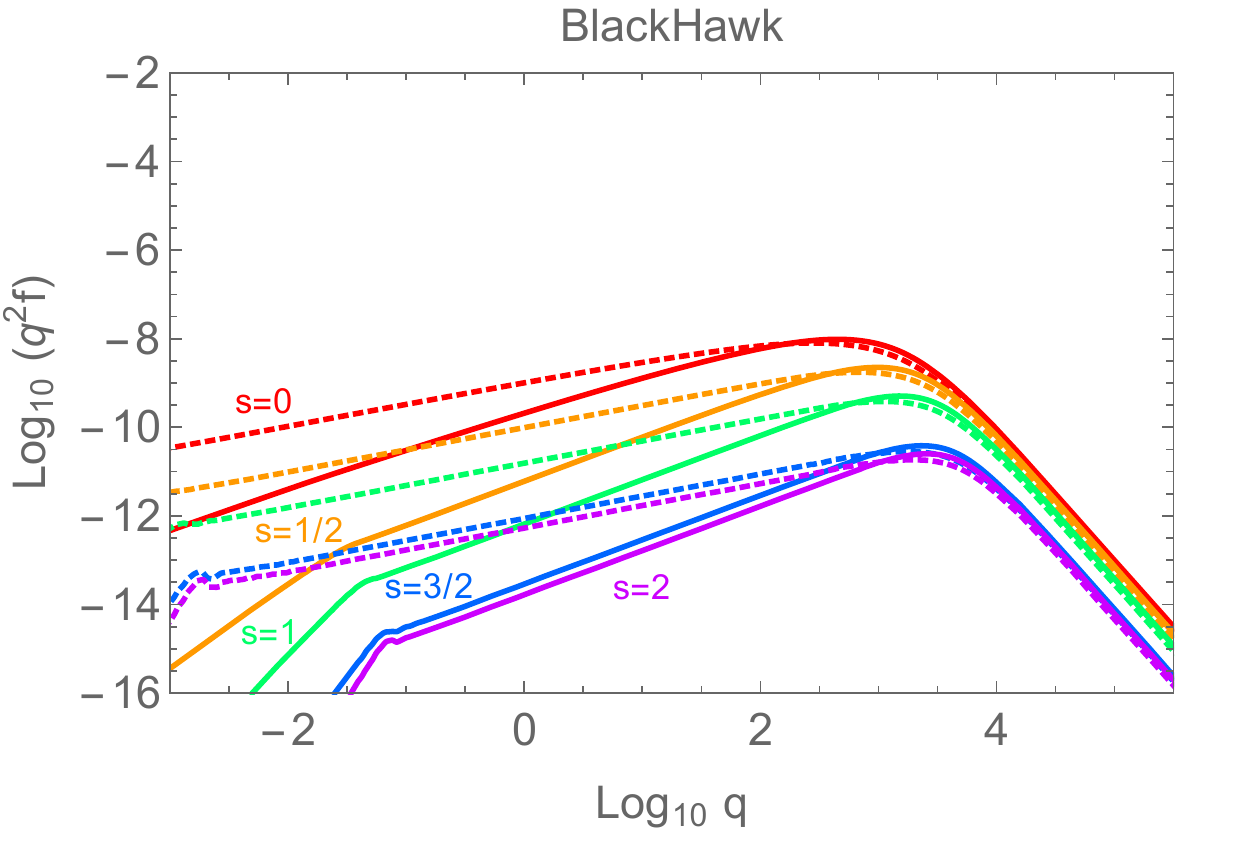}
	\includegraphics[width = 0.49\textwidth]{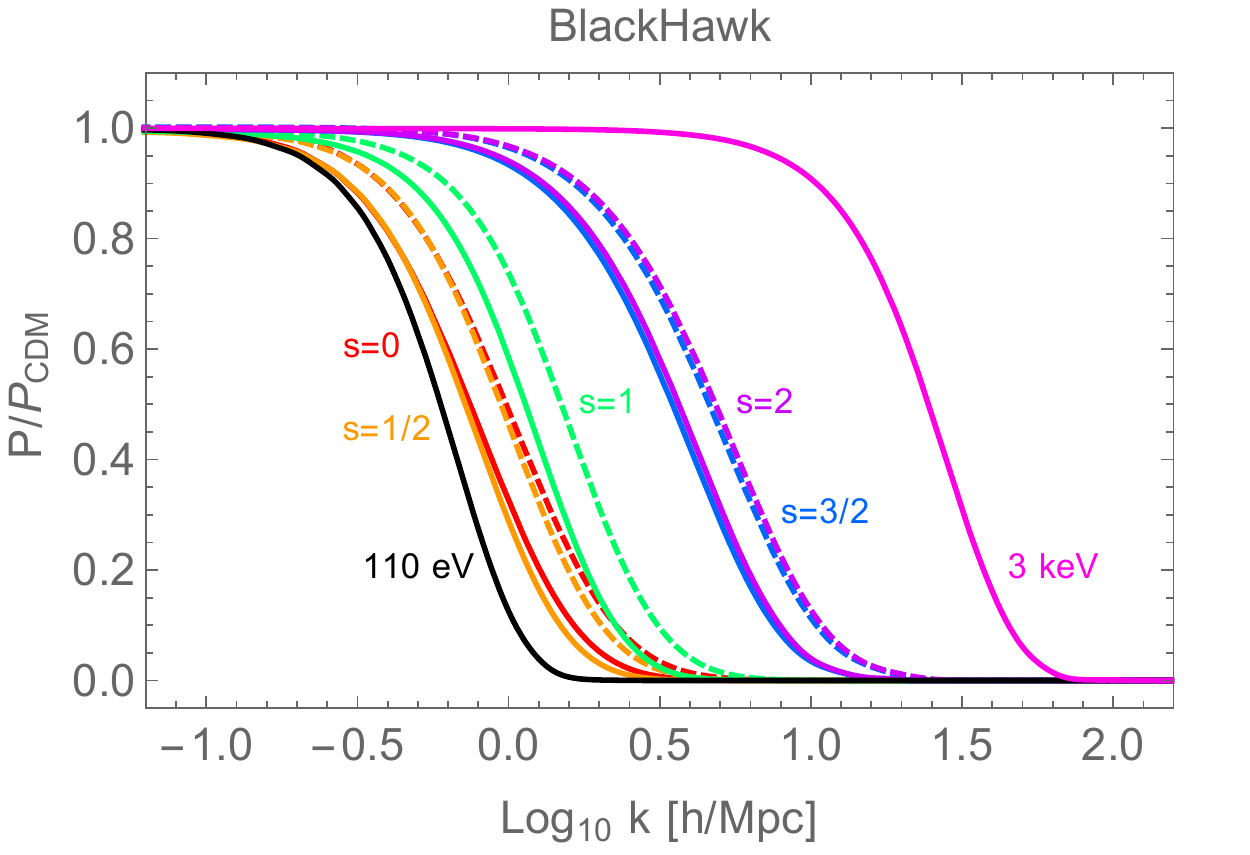}
	\caption{\textbf{Left:} Phase space distribution of different spin DM evaporated by a PBH, at final evaporation time. \textbf{Right:} Transfer function of the matter power spectrum at CMB epoch, computed by \texttt{CLASS}. [taken from~\cite{Auffinger:2020afu}]}
	\label{fig:BSM}
\end{figure}

Dark sector emission was included inside \texttt{BlackHawk v2.0} by modifying the number of degrees of freedom (dofs) that a BH emits as a function of its temperature $T$. The formula describing the mass loss of a BH is
\begin{equation}
	\dfrac{\d M}{\d t} = \dfrac{-f(M)}{M^2}\,,
\end{equation}
where $f(M)$ is the Page coefficient that counts the available dofs. Increasing $f(M)$ makes the BH evaporate faster. The spectrum of dark sector particles emitted is computed together with the SM ones.

As an application, one can compute the amount of DM particles evaporated from PBHs in the early universe, in a tentative to conciliate the origin of DM with its absence of interaction with the SM particles. Indeed, PBH evaporation creates all kinds of particles regardless of their interactions. This work was performed in \cite{Auffinger:2020afu}, improving the results of \cite{Masina:2020xhk} thanks to the new \texttt{BlackHawk} version. In this study, we computed the amount of DM radiated by PBHs in the lowest possible mass range $10^{-1}-10^{9}\,$g, unconstrained because these PBHs evaporate before the BBN. We obtained in particular the phase space distribution (PSD) of DM at the end of the PBH lifetime $\tau \propto M^3$ by integrating over the \texttt{BlackHawk} spectrum (see Fig.~\ref{fig:BSM}, left panel). Then, we used the public code \texttt{CLASS}~\cite{Lesgourgues:2011re,Lesgourgues:2011rh} to obtain the transfer function of the power spectrum at the CMB epoch (see Fig.~\ref{fig:BSM}, right panel). This transfer function is constrained in the following way: if the DM particles are too hot at CMB times, then they could spoil small scale structure formation. This results in a constraint on the DM PSD, and consequently on the DM mass $m$, PBH temperature at evaporation $T \propto 1/M$ and equivalently PBH lifetime $\tau \propto M^3$. As a conclusion, we obtained constraints on the abundance of PBHs $\beta$ in the disputed mass range $M \lesssim 10^9\,$g, which depend on the DM particle spin $s$ since the HR emission rates are strongly spin-dependent. This study was extended to spinning PBHs in~\cite{Masina:2021zpu}, still using \texttt{BlackHawk}.

\subsection{Spin 3/2 greybody factors}
\label{sec:32}

\begin{figure}
	\centering
	\includegraphics[width = \textwidth]{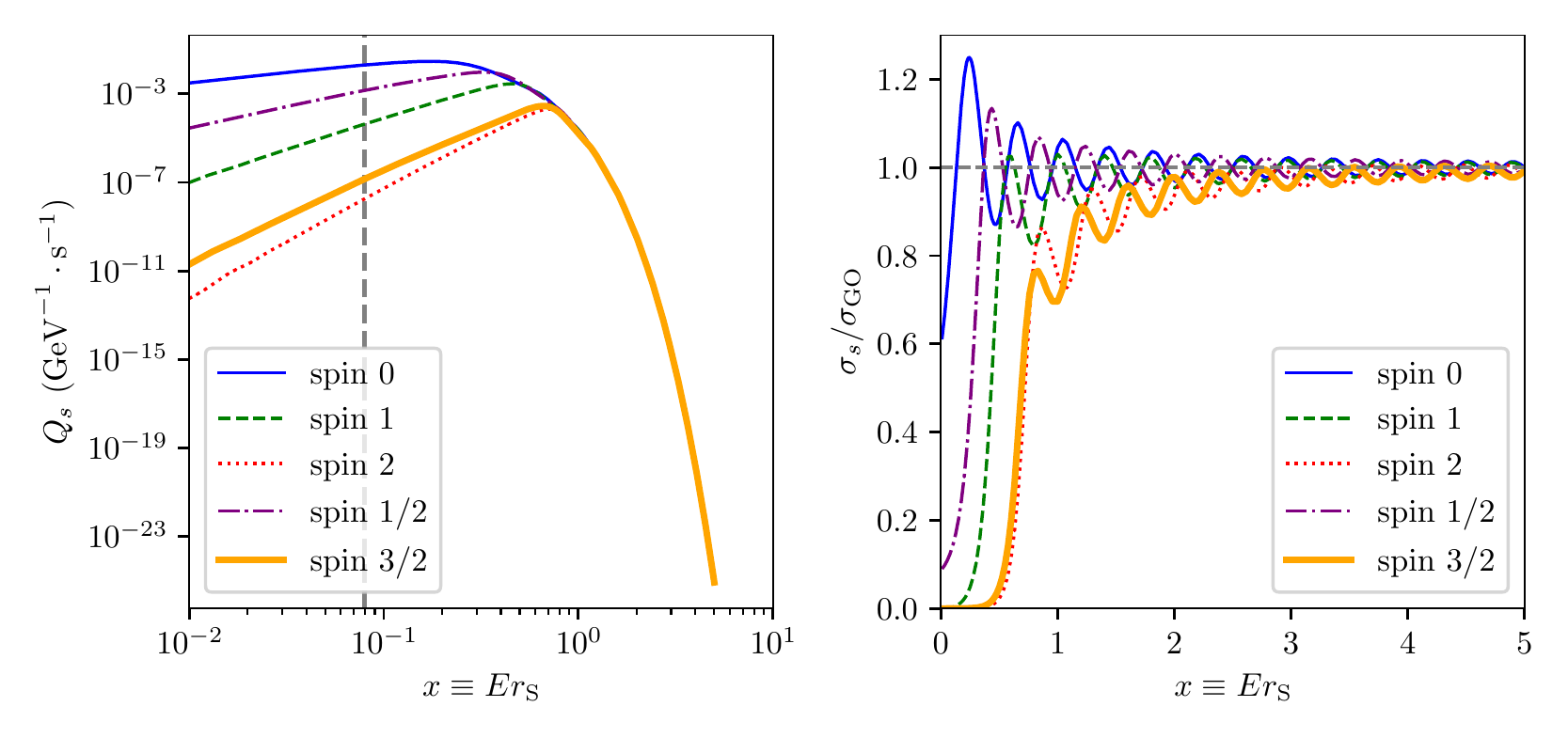}
	\caption{\textbf{Left:} Hawking emission rate \eqref{eq:Hawking} of a single dof of different spins. \textbf{Right:} The corresponding cross-sections. \textit{Emphasis is put on the spin 3/2 results.} [taken from~\cite{Arbey:2021mbl}]}
	\label{fig:32}
\end{figure}

The BSM study described above calls for non-standard particle emission, in particular spin $3/2$ particles that are natural components of the supersymmetric set of particles. The gravitino is a perfect example. However, the greybody factors $\Gamma$ of spin $3/2$ particles required to compute the HR rate of a type $i$ field\cite{Hawking:1974rv,Hawking:1975vcx}
\begin{equation}
	\dfrac{\d^2 N_i}{\d t\d E} = \dfrac{1}{2\pi} \dfrac{\Gamma}{e^{E/T} - (-1)^s}\,,\label{eq:Hawking}
\end{equation}
were absent from the literature, and only approximations were used when considering gravitino emission by PBHs~\cite{Khlopov:2004tn}. Inside \texttt{BlackHawk v2.0}, we followed the method of Chandrasekhar and Detweiler to obtain numerically the greybody factors by solving the Teukolsky equation (see the book ``Mathematical theory of black holes'' from Chandrasekhar for a detailed account of this method~\cite{Chandrasekhar:1985kt}). The result is shown in Fig.~\ref{fig:32} together with the other well-known spin 0, 1, 2 and 1/2 cross-sections for a Schwarzschild BH.

\subsection{Non-standard black hole metrics}
\label{sec:BSMmetrics}

\begin{figure}
	\centering
	\includegraphics[width = 0.6\textwidth]{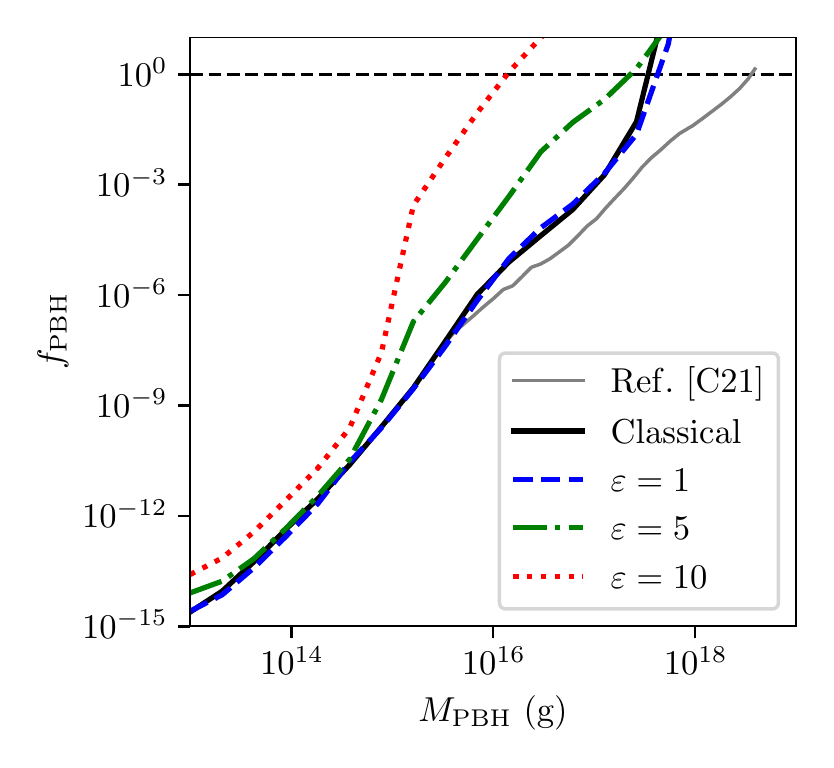}
	\caption{AMEGO constraints on polymerized PBHs, for different value of the polymerization parameter $\varepsilon$, compared to the Schwarzschild results from [C21]~\cite{Coogan:2020tuf}. [taken from~\cite{Arbey:2021yke}]}
	\label{fig:LQG}
\end{figure}

The PBH constraints from HR have been refined in an impressive number of ways in just a few years (see the reviews~\cite{Carr:2020gox,Green:2020jor}). However, most of the constraints rely on the assumption that PBHs are of the Schwarzschild type, with some extension to include Kerr rotating PBHs in the recent literature. Non-standard BH solutions, even if they are thoroughly studied from a theoretical point of view, are not often used to obtain constraints on PBHs.\footnote{A remarkable exception is the case of higher-dimensional BHs.} In the two companion papers \cite{Arbey:2021jif,Arbey:2021yke}, we performed a general study of the spherically-symmetric and static metrics and obtained the Teukolsky equation. Solving this Teukolsky equation for a particular metric then raises the greybody factor values that are used inside \texttt{BlackHawk} to obtain the HR rate (see Eq.~\eqref{eq:Hawking}). As an application, we computed the PBH constraints from the prospective AMEGO in \cite{Arbey:2021yke} for both a Schwarzschild and a polymerized BH (see Fig.~\ref{fig:LQG}), the latter arising in effective extensions of the loop quantum gravity (LQG) theory. It is particularly interesting to see that the PBH constraints are modified in the case of non-standard PBH metrics, in particular for regular metrics that do not exhibit a coordinate singularity, such as the polymerized BH. The general method described in \cite{Arbey:2021jif,Arbey:2021yke} can be extended to a wide family of BH (regular) solutions.

\subsection{Low- and high-energy hadronization}
\label{sec:Hazma}

\begin{figure}
	\centering
	\includegraphics[width = 0.49\textwidth]{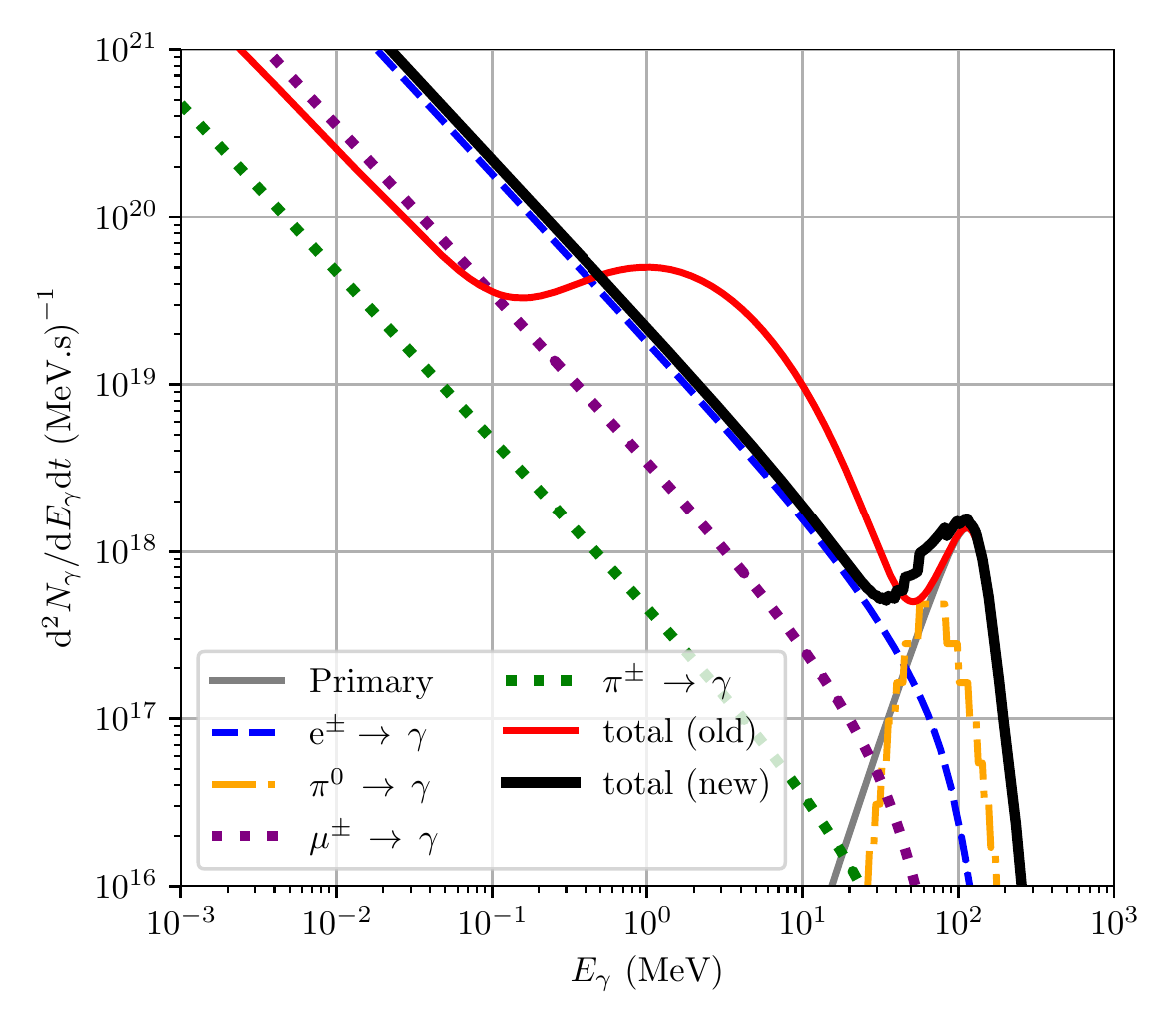}
	\includegraphics[width = 0.49\textwidth]{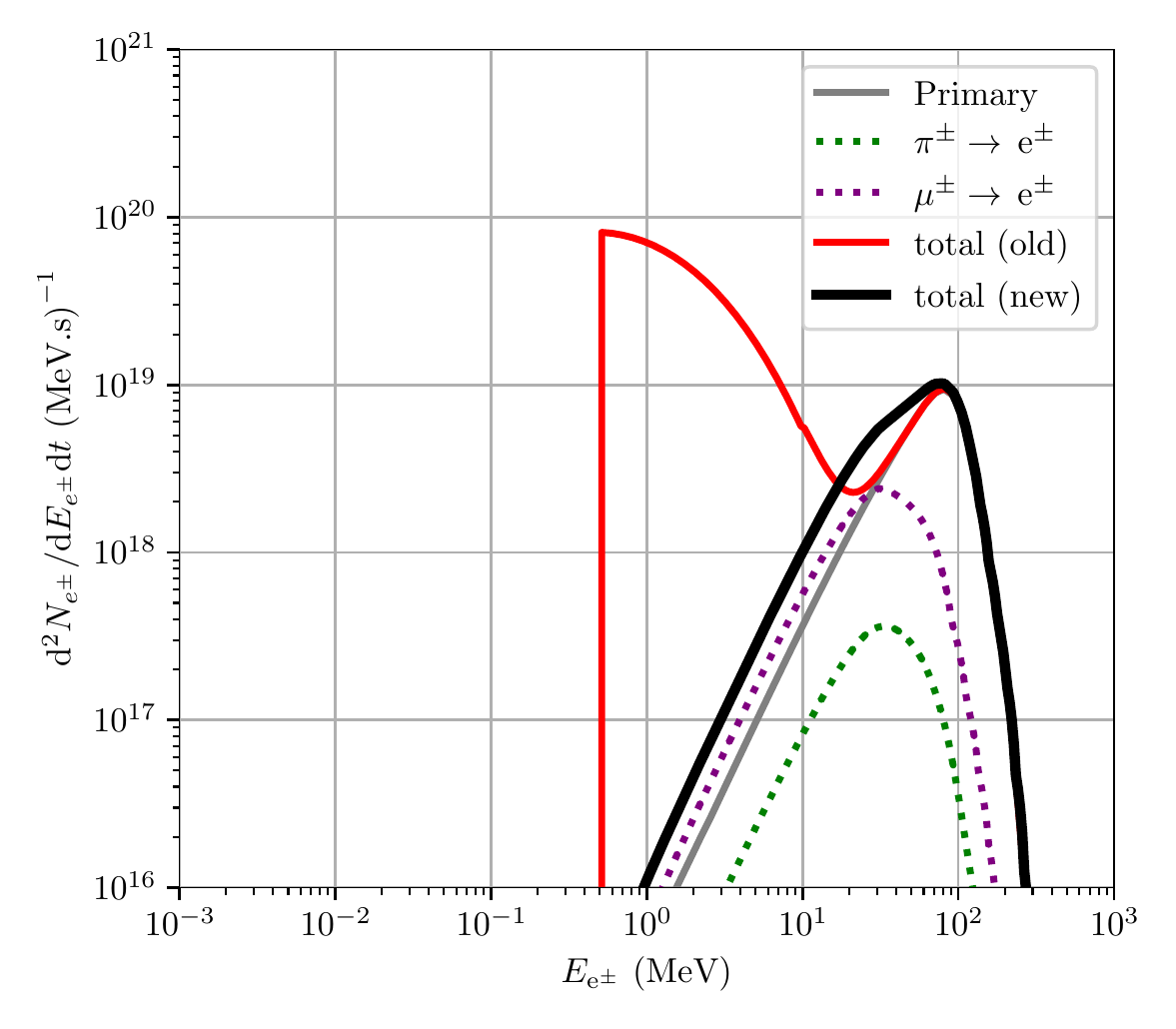}
	\caption{\textbf{Left:} Photon secondary spectra of a $5.3\times 10^{14}\,$g BH computed using the \texttt{Hazma} package, with the contribution of the different channels. \textbf{Right:} The corresponding electron secondary spectra. [taken from~\cite{Arbey:2021mbl}]}
	\label{fig:lowE}
\end{figure}

The AMEGO constraint discussed above raised an issue: PBHs in the mass range just above the evaporation limit $5\times 10^{14}\,$g have a temperature in the low energy QCD domain $E \sim 100\,$MeV. This represented a difficulty for \texttt{BlackHawk v1.2}, which relied on \texttt{PYTHIA}~\cite{Bierlich:2022pfr} and \texttt{HERWIG}~\cite{Bellm:2019zci} to compute the hadronization of primary particles, valid for $E\gtrsim$ GeV. There was a need for more careful treatment of this energy range, a work initiated by the study of Ref.~\cite{Coogan:2020tuf} based on the public package \texttt{Hazma}~\cite{Coogan:2019qpu}. We interfaced \texttt{BlackHawk} with \texttt{Hazma} to incorporate realistic hadronization tables that allow to go from the HR primary spectrum of Eq.~\eqref{eq:Hawking} to the final stable output of a BH
\begin{equation}
	\dfrac{\d^2 \tilde{N}_j}{\d t\d E} = \sum_i \int_0^{+\infty} \text{Br}_{i\rightarrow j}(E,E^\prime) \dfrac{\d^2 N_i}{\d t\d E^\prime}\d E^\prime\,,
\end{equation}
where $\text{Br}_{i\rightarrow j}$ are the branching ratios. \texttt{Hazma} provides the branching ratios for $j = $ photons and electrons, lacking for now the neutrino outcome. The secondary spectra are shown in Fig.~\ref{fig:lowE}. Hadronization tables have also recently been obtained in the ultra-high energy regime $E \gg $ TeV thanks to the \texttt{HDMSpectra}~\cite{Bauer:2020jay} package, based on Ref.~\cite{Capanema:2021hnm}.

\section{Conclusion}

Since the preceding TOOLS2020 conference~\cite{Auffinger:2020ztk}, \texttt{BlackHawk} has received much upgrades, including the possibility of BSM particle emission, HR from non-standard BH solutions and a careful treatment of hadronization at low- and high-energy. \texttt{BlackHawk} continues to be updated to incorporate new features, the last to date being the \texttt{Isatis} tool that computes automatically the constraints on PBHs from HR with a wide range of instruments~\cite{Auffinger:2022dic}. \texttt{Isatis} is a powerful tool to examine the precise dependency of the constraints on the HR theory and the instrumental characteristics. \textit{The \texttt{BlackHawk} authors are available by email to answer any query you could have about the code.}

\bibliographystyle{JHEP}
\bibliography{biblio.bib}

\end{document}